\documentclass[a4paper,11pt]{article}
\usepackage{pos}

\title{The chiral critical point from the strong coupling expansion}
\author[a]{Jangho Kim}
\author[a]{Pratitee Pattanaik}
\author*[a]{Wolfgang Unger}

\affiliation[a]{Fakult\"at f\"ur Physik, Universit\"at Bielefeld,\\
Universit\"atstrasse 25, D33619 Bielefeld, Germany}

\emailAdd{wunger@physik.uni-bielefeld.de}
\emailAdd{pratiteep@physik.uni-bielefeld.de}
\emailAdd{jangho@physik.uni-bielefeld.de}

\newcommand{\nn}{\nonumber \\}
\newcommand{\lr}[1]{\left( #1 \right)}

\newcommand{\Nt}{{N_{\tau}}}

\newcommand{\Ocal}{\mathcal{O}}    % e.g. Observable, Order

\newcommand{\Nf}{{N_{\rm f}}}

\newcommand{\Nc}{{N_{\rm c}}}
\newcommand{\Ord}[1]{\Ocal(#1)}
\newcommand{\SU}{{\rm SU}}
\newcommand{\U}{{\rm U}}

\abstract{The strong coupling expansion for staggered fermions allows for Monte Carlo simulations using a dual representation. It has a mild sign problem for low values of the inverse gauge coupling 
$\beta$, hence the phase diagram in the full $\mu_B - T$ plane can be evaluated. We have extended this framework to include 
$\Ord{\beta}$ and $\Ord{\beta^2}$ corrections, by mapping the degrees of freedom to a vertex model. We present results on the $\beta$-dependence of the chiral critical point from those simulations.
}

\FullConference{The 41st International Symposium on Lattice Field Theory (LATTICE2024)\\
 28 July - 3 August 2024\\
Liverpool, UK\\}

\begin{document}
\maketitle

\section{Introduction}

Lattice QCD at finite density suffers from the notorious sign problem that prevents to determine the QCD phase diagram in the $\mu_B - T$ plane. Despite several recent developments in lattice QCD at finite density \cite{Borsanyi:2021,Bollweg:2022,Clarke:2024,Hansen:2024}, the search for possible QCD critial point has not yet been conclusive.  One of several strategies to address the phase diagram in a regime of lattice QCD where the sign problem is manageable is to make use of a representation with a much milder sign problem, based on color singlet states. Those color singlets naturally arise in the so-called dual representation, where gauge links have been integrated out systematically in a strong coupling expansion (SCE) in the inverse gauge coupling $\beta=\frac{2N_c}{g^2}$.

The dual representation in terms of a monomer-dimer-polymer system is well established since decades in the strong coupling limit $\beta=0$ for staggered fermions  \cite{Rossi:1984}. The key insight was to reverse the order of integration, i.e.~integrating out the gauge links $U_\mu(x)$ first, then the Grassmann-valued quarks. Hence there is no fermion determinant $\det D[U]$, instead the partition function is in terms of integer variables $\{m_x, k_\ell, b_\ell$\} , with the link coordinate $\ell=(x,\mu)$, which can be sampled via Monte Carlo.
More recently, it has been extended to higher order in $\beta$, based on a Taylor expansion of the Wilson gauge action \cite{Gagliardi:2019}. The standard dual variables in the strong coupling limit are (1) the monomers $m_x\in \{0,\ldots \Nc\}$ that encode chiral observables and which introduce the quark mass dependence, (2) the dimers $k_\ell\in \{0,\ldots \Nc\}$ which encode pion propagation, and (3) a set of non-intersecting loops that encode baryon propagation $b_\ell\in\{\pm 1\}$ and introduce the dependence on the baryon chemical potential. Temperature is introduced via an anisotropy in temporal direction, affecting  temporal meson and baryon hops. 
The Taylor expansion in $\beta$ introduces three types of additional dual variables: (4) (anti-) plaquette occupation numbers $n_p$, $\bar{n}_p$, (5) quark propagation $d_\ell\in\{\pm 1,\pm 2, \pm3\}$ on the links and (6) multi-indices $\rho_\ell$ that are termed decoupling operator indices for reasons discussed in the next section. We will shortly review that this gives rise to a tensor network representation of lattice QCD.

The first study to include leading order gauge corrections to the strong coupling limit used reweighting in the plaquette occupation number \cite{deForcrand:2014} to address the $\mu_B - T$ phase diagram. Actual simulations at  $\mathcal{O}(\beta)$, sampling the plaquette occupation numbers $(n_p,\bar{n}_p)\in\{(1,0),(0,1)\}$ with an additional update were addressed in \cite{Kim:2023} concerning the nuclear liquid gas transition. Here we want to include the $\mathcal{O}(\beta^2)$ corrections, sampling also the $(n_p,\bar{n}_p)\in\{(2,0),(1,1),(0,2)\}$ contributions to the partition function to determine the effects on the phase diagram. Although the strong coupling expansion is truncated at a rather low order, the strong coupling regime of lattice QCD shares important features with full QCD, such as confinement, chiral symmetry breaking and its restoration at finite temperature, and a nuclear transition from the vacuum to a phase with high baryon density at low temperatures.
Another benefit is that fast simulations are possible via the worm algorithm, hence no supercomputers are necessary and the scan of the various parameters can be performed on a compute clusters. However, at $\Ord{\beta^2}$, simulations slow down drastically.

In this work we focus on the phase diagram for $\Nf=1$ in the chiral limit. The critical point that exists at strong coupling at finite quark mass becomes a tri-critical point in the chiral limit \cite{Kim:2016}. 
Results for $\Nf=2$ in the strong coupling limit have also been obtained, but not yet extended to low temperatures to pin-point the critical point \cite{Unger:2024}.

\section{From tensor formulation to the simulation of the vertex model}

We use a ``standard'' QCD lattice action: one flavor of unrooted staggered fermions and the Wilson gauge action, both unimproved to facilitate gauge integration. As in the strong coupling limit, i.e. without the inclusion of the Wilson gauge action, the lattice is maximally coarse, it is important to include gauge corrections from the expansion of the gauge action 
to make the lattice finer. For $\beta>0$, the strong coupling expansion also gives rise to a dual representation in terms of colour singlets, including the above mentioned additional degrees of freedom. Furthermore it has been mapped to a a tensor network formulation \cite{Gagliardi:2019} that we shorty review:

\newcommand{\D}{{\mathcal{D}_b}}
\newcommand{\E}{{\mathcal{E}_{a-b}}}
\newcommand{\DD}[1]{{\mathcal{D}_{#1}}}
\newcommand{\EE}[1]{{\mathcal{E}_{#1}}}
\newcommand{\tWg}{{\rm \tilde{W}\hspace{-0.7mm}g}}
\makeatletter
\newcommand{\raisemath}[1]{\mathpalette{\raisem@th{#1}}}
\newcommand{\raisem@th}[3]{\raisebox{#1}{$#2#3$}}
\newcommand{\ijkl}{_{i\,\raisemath{3pt}{j},k\,\raisemath{3pt}{l}}}

The starting point is a combined Taylor expansion in the reduced gauge coupling $\tilde{\beta} \equiv \frac{\beta}{2N}=\frac{1}{g^{2}}$  and quark mass $\hat{m}_{q}$, 
 giving rise to the dual variables: $n_p$, $\bar{n}_p$, $d_\ell$, $\bar{d}_\ell$ and $m_x$:
 \begin{align}
\mathcal{Z}(\beta,\mu_q,\hat{m}_q)&=
{\displaystyle \sum_{\substack{\{n_{p},\bar{n}_{p}\} \\ \{d_{\ell},\bar{d}_{\ell},m_{x} \}  }}}\!\!\!\prod_{p}\frac{\tilde{\beta}^{n_{p}+\bar{n}_{p}}}{n_{p}!\bar{n}_{p}!}\prod_{\ell}\frac{1}{d_{\ell}!\bar{d}_{\ell}!}{\displaystyle \prod_{x}}\frac{(2\hat{m}_{q})^{m_{x}}}{m_{x}!}\,\boldsymbol{\mathcal{G}}_{n_{p},\bar{n}_{p},d_{\ell},\bar{d}_{\ell},m_{x}}.
\end{align}
The evaluation of the 1-link integrals 
\begin{align}
\label{I-integral-computation}
\mathcal{I}^{a,b}\ijkl = \int_{\SU(N)}DU\,\, U_{i_{1}}^{\,j_{1}} \cdots U_{i_{a}}^{\,j_{a}}U^{\dag \,l_{1}}_{k_{1}} \cdots U^{\dag \,l_{b}}_{k_{b}}
 &\propto\!\!\!{\displaystyle \sum_{(\alpha,\beta)}}{\displaystyle \sum_{\pi,\sigma \in S_{p}}}\epsilon^{\otimes q}_{i_{\{\alpha\}}}\delta_{i_{\{\beta\}}}^{l_{\pi}}\tWg_{N}^{q,p}(\pi\circ\sigma^{-1})\epsilon^{\otimes q, j_{\{\alpha\}}}\delta_{k_{\sigma}}^{\;j_{\{\beta\}}}.
\end{align}
within $\boldsymbol{\mathcal{G}}$ is in terms of generalized Weingarten functions, $\tWg_{N}^{q,p}$ i.e. the $\SU(N)$ generalization of the usual $\U(N)$ Weingarten functions. 
Here, the color indices $i=i_1,\ldots i_a$, $j=j_1,\ldots j_a$, $k=k_1,\ldots k_b$, $l=l_1,\ldots l_b$ enter the epsilon tensors and Kronecker deltas; note that $p=\min(a,b)$, $q=|a-b|/\Nc$.
We decouple those integrals via a choice of orthogonal projector operators $P^\rho$ into a local tensor 
$\mathcal{T}_{x}^{\rho^{x}_{-d}\cdots\rho^{x}_{d}}$ that depends on participating dual degrees of freedom: $\mathcal{D}_{x}=\left\{m_{x},d_{x,\pm\mu}, n_{x,\mu\nu},\bar{n}_{x,\mu\nu}\right\}$.
For each (positive and negative) direction $\pm d$ from site $x$, the decoupling operator indices (DOI) $\rho=[\alpha,\beta,(m,n)_\lambda]$ are multi-indices which depend on the set of parameters $\alpha$, $\beta$ which enter the product of epsilon tensor and Kronecker deltas (forming the gauge invariant basis), and on the irreducible representation $\lambda$. 
The final dual partition function is:
\begin{align}
\mathcal{Z}(\beta,\mu_{q},\hat{m}_q) = \hspace{-3mm}{\displaystyle \sum_{\substack{\{n_{p},\bar{n}_{p}\} \\ \{k_{\ell},f_{\ell},m_{x} \}  }}}\hspace{-2mm}
\sigma_{f} {\displaystyle \sum_{\{\rho^{x}_{\pm\mu}\}}}\!\!\!\prod_{p}\frac{\tilde{\beta}^{n_{p}+\bar{n}_{p}}}{n_{p}!\bar{n}_{p}!}\prod_{\ell=(x,\mu)}\frac{e^{\mu_{q}\delta_{\mu,0}f_{x,\mu}}}{k_{\ell}!(k_{\ell}+|f_{\ell}|)!}{\displaystyle \prod_{x}}\frac{(2\hat{m}_q)^{m_{x}}}{m_{x}!}\mathcal{T}_{x}^{\rho^{x}_{-d},\dots,\rho_{d}^{x}}(\mathcal{D}_{x})
\end{align}
with the dimer $k_\ell=\min(d_\ell,\bar{d}_\ell)$ and the residual fermion flux $f_\ell=d_\ell-\bar{d}_\ell$. The number of distinct DOI depends on the order of the SCE, at $\Ord{\beta^2}$, there are 6 distinct ones. They can be cast into integers and thus the tensors can be cast into vertices.
In contrast to studies at strong coupling using tensor network approaches via singular value decomposition (so war limited to two dimensions) \cite{Bloch:2022}, in order to address 3+1 dimensional lattices, we do not consider tensor contractions, but use a corresponding vertex model for which we devise a Monte Carlo algorithm.

Vertex models are a type of statistical models where the Boltzmann weight is assigned to a vertex and cannot be decomposed into site and bond weights. 
In the context of ferroelectric crystals (ice type vertex models), the weights for different incoming/outgoing directions reflect physical constraints or conservation laws. 
In the context of field theory, the 8 vertex model for the Schwinger model with Wilson fermions in the strong coupling limit has been studied \cite{Wenger:2008}.
Our vertex model has a much higher number of vertices $N_v$, as summarized in Tab.~\ref{Tab:Vertices}. Each bond state $v_b$ is a combination of the DOI, the fermion and anti-fermion flux $d_l$, $\bar{d}_l$ and the plaquette occupation numbers participating in a given bond $b$. Each vertex $v$ has thus the structure $v=(v_0,\ldots v_{2D})$ with $D$ the dimension of the lattice ($D=4$) in our case. There are $N_v$ distinct vertices $v^{(i)}, i=1,\ldots N_v$ and the weight $w(v^{i})$ which also contains the physical parameters quark mass $am_q$, anisotropy $\gamma$, quark chemical potential $\mu$ and inverse gauge coupling $\beta$ are tabularized prior to the Monte Carlo simulation and is a product of the weight based on dual variables $\mathcal{D}$ and the tensor $\mathcal{T}$:
\begin{align}
 w(v^{(i)})&=w_{\mathcal{D}}(v^{(i)})w_{\mathcal{T}}(v^{(i)}),\nn
 w_{\mathcal{D}}(v^{(i)})&=
 (2am_q)^{m_x} 
 \gamma^{(k_{x,t}+k_{x,-t}+|f_{x,t}|+|f_{x,-t}|)/2} 
 e^{a\mu_B (f_{x,t}+f_{x,-t})/(2\gamma^2) }
 \lr{\frac{\beta}{2\Nc}}^{(n_p+\bar{n}_p)/4},
\end{align}
where bonds and plaquettes weights in $w_{\mathcal{D}}(v^{(i)})$ are equally distributed with the neighboring vertices they connect to.

\begin{figure}[!ht]
\centerline{\includegraphics[width=\textwidth]{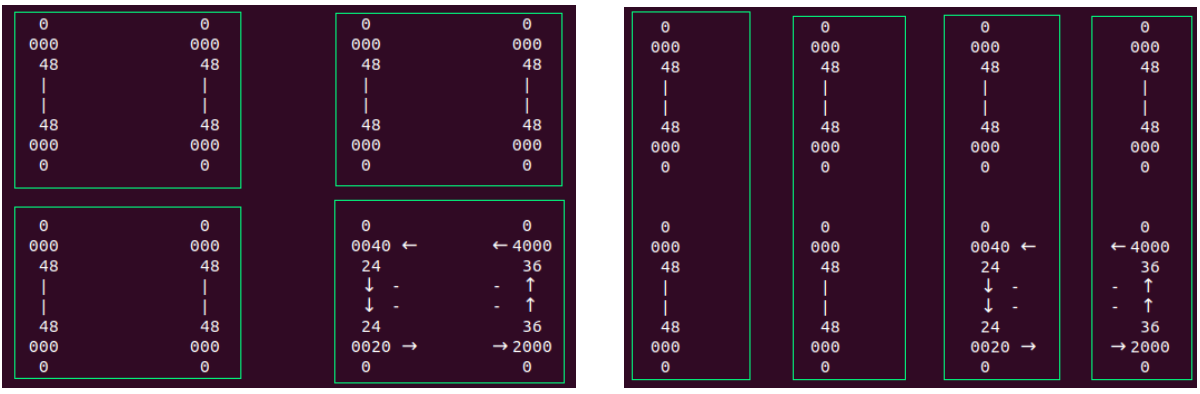}}
\caption{Vertex updates (dimers and quark fluxes are plotted for illustration). \emph{Left:} vertices updated along elementary plaquettes. \emph{Right:} vertices updated along static lines.}
\label{Fig:Update}
\end{figure}

As the state space is discrete, we use a heatbath algorithm for this vertex model to update the vertex configurations. In Fig.~\ref{Fig:Update} we show two types of updates that modify the vertices on closed contours: the plaquette update and the line update, which in conjunction gives an ergodic algorithm \cite{Pattanaik:2023}. It is also evident that both updates can be parallelized, which we did to speed up the simulations significantly. It should be noted that in contrast to the strong coupling limit, where a worm algorithm is the best choice, the non-trivial DOI hinders a straight-forward application of the worm algorithm. All results at $\Ord{\beta}$ and $\Ord{\beta^2}$ have been compared to exact enumeration in small volumes and crosschecked with hybrid Monte Carlo for $\mu_B=0$. Hence we expect our simulations at non-zero $\mu_B$ to be accurate.
The lattices for which we gathered simulations are the volumes $8^3\times4$, $12^3\times4$ and $16^3\times 4$, in the strong coupling limit and for  $\mathcal{O}(\beta)$, $\mathcal{O}(\beta^2)$ in the range $\beta=[0.0,\ldots, 1.0]$ and at temperatures $T=0.8,0.85,0.9,0.95,1.0$, all in the chiral limit.

\begin{table}
\begin{center}
\begin{tabular}{llrrrr}
 \hline
 %dimension &  
 limit & $\mathcal{O}(\beta^0)$ & $\mathcal{O}(\beta^1)$ & $\mathcal{O}(\beta^2)$ & $\mathcal{O}(\beta^3)$\\
 \hline
% $d=2$:& all  & 47 & 255 & 1499 & 8939\\
% $d=2$:& chiral & 32 & 192 & 1176 & 7264\\
% $d=2$:& quenched & 1 & 1 & 5 & 13\\
% \hline
 %$d=4$:& 
 all & 221 & 3485 & 51125 & 681013\\
 %$d=4$:& 
 chiral & 176 & 2960 & {\color{red}44672} & 607792\\
 %$d=4$:& 
 quenched & 1 & 1 & 25 & 137\\
 \hline
\end{tabular}
\end{center}
\caption{Number of distinct vertices depending on the order of the strong coupling expansion, highlighted in red the vertices we used in this work}
\label{Tab:Vertices}
\end{table}

%\begin{figure}
%\centerline{\includegraphics[width=0.45\textwidth]{DecoupOp.png}\includegraphics[width=0.45\textwidth]{DecoupOp.png}}
%\caption{Left: tensor composed of dual variables, including decoupling operator indices. Right: vertices that also contain %the modified weights required for the Monte Carlo simulations.}
%\end{figure}

\section{The sign problem at finite $\beta$}

\begin{figure}[!ht]
\centerline{\includegraphics[width=1.1  \textwidth]{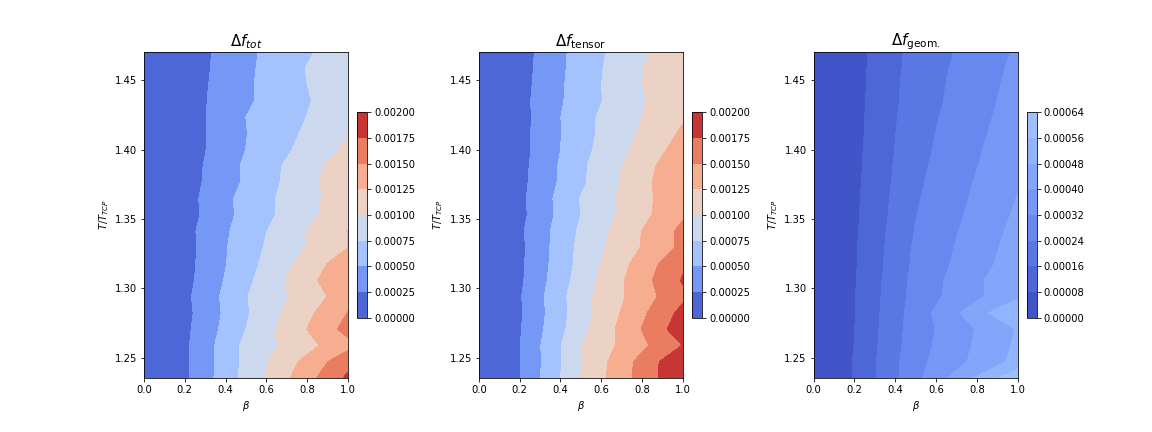}}
\caption{Decomposition of the total sign \emph{(left)} into tensor/vertex \emph{(center)} sign and geometric sign \emph{(right)}, using $\Delta f$ and for $\Ord{\beta^2}$. The tensor sign dominates the total sign in the whole range of parameters.}
\label{Fig:Sign}
\end{figure}

As pointed out in the introduction, the reason to resort to the strong coupling expansion by means of the dual representation is that the residual sign problem is much milder compared to conventional Monte Carlo based on the fermion determinant. Hence, sign reweighting is feasible for rather large volumes to study the phase transitions at finite baryon chemical potential. 
The reason for this mild sign problem of order $\Delta f\simeq 10^{-5}$ in the vicinity of the tri-critical point is that the sign is purely geometrical and fermionic in nature: at strong coupling, it only depends on the baryonic world-lines:
Only few loops that have non-trivial geometries contribute with a minus sign (originating from Fermi statistics, anti-periodic boundary conditions, backward hops, and product of staggered phases along the loop), while most loops are positive. This can be understood by the fact that baryons are heavy, hence non-relativistic, and rarely hop in spatial directions, as compared to pions or individual quarks. Also, the color singlets are closer to the physical states than the gauge configurations one has to deal with in the standard methods. Moreover, as the gauge links $\U_\mu(x)$ have been integrated out, no fluctuations from them enter the sign problem. It is worth noting that the sign problem is maximal in the chiral limit and at low temperatures, and gets reduced with increasing quark mass or temperature \cite{Kim:2023}, which result in simpler geometries of the baryonic loops.

However, based on the value of $\beta$, the sign problem gets re-introduced as color singlets are now composed of quark and gluon contributions, and the sign problem is only manageable in the regime $\beta=\frac{6}{g^2} \lesssim 1$ if based on the Taylor expansion. In Fig.~\ref{Fig:Sign} we show the nature of the full sign problem, which can be composed into the geometric sign and the tensor sign, which has a dominant contribution at $\Ord{\beta^2}$. Note that the tensor sign is not present in the strong coupling limit. 
We will see in the next section, that due to the severity of the sign problem at $\beta=1$, finite size scaling is not yet feasible due to the limited volumes to take into account, in particular for the $\Ord{\beta^2}$ simulations.

It is desirable to resort to a character expansion instead, which would push the range of validity from 
$\beta\lesssim 1$ to $\beta\lesssim 2\Nc$ with the resummation involved on the fermion lines. However, in contrast to Wilson fermions where the character expansion is well established for heavy-dense lattice QCD \cite{Langelage:2014}, staggered fermions allow for fermion back-tracking (giving e.g.~rise to dimers), which hinders a straight-forward resummation into characters. 
Whether it is feasible at all remains an open question.

\section{Results in the vicinity of the tri-critical point}

\begin{figure}[!htb]
\includegraphics[width=\textwidth]{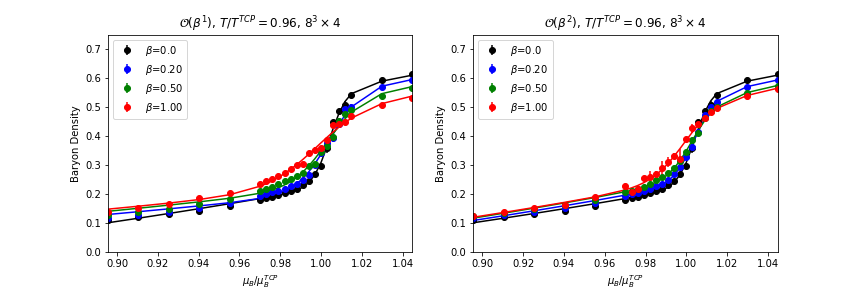}
\includegraphics[width=\textwidth]{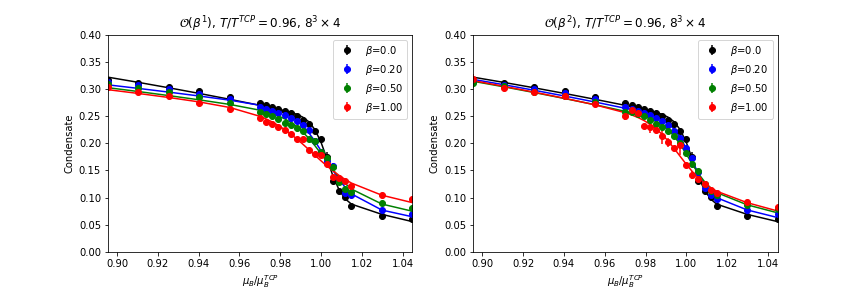}
\caption{\emph{Top:} baryon density, \emph{bottom:} chiral condensate; MC simulations for at $\Ord{\beta}$ \emph{(left)} or $\Ord{\beta^2}$ \emph{(right)}.}
\label{Fig:Density}
\end{figure}

\begin{figure}[!htb]
\includegraphics[width=\textwidth]{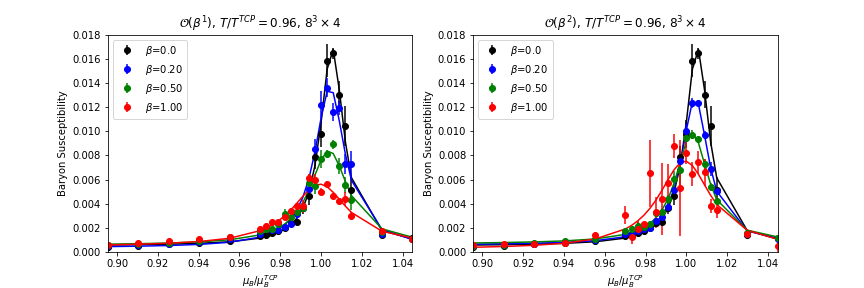}
\includegraphics[width=\textwidth]{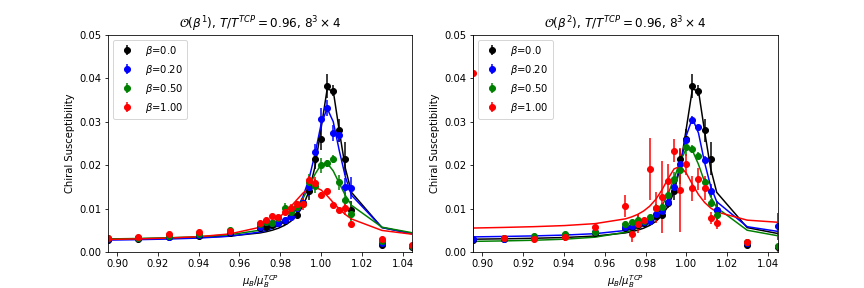}
\caption{\emph{Top:} baryon susceptibility, \emph{bottom:} chiral susceptibility; MC simulations for at $\Ord{\beta}$ \emph{(left)} or $\Ord{\beta^2}$ \emph{(right)}.}
\label{Fig:Susceptibility}
\end{figure}

\begin{figure}[!htb]
\includegraphics[width=\textwidth]{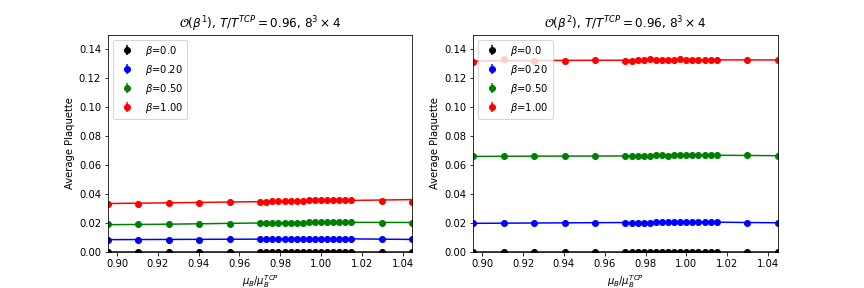}
\caption{Average plaquette from MC simulations for at $\Ord{\beta}$ \emph{(left)} or $\Ord{\beta^2}$ \emph{(right)}.}
\label{Fig:Plaquette}
\end{figure}

 All results shown in the figures are relative to the location of the strong coupling chiral tri-critical point (TCP), which is located at $T^{TCP}_{\Nt=4}=0.83$ and ${\mu_B}^{TCP}_{\Nt=4}=2.07$ \cite{Kim:2016}.
In Fig.~\ref{Fig:Density} we show both the baryon density and chiral condensate as fixed temperature just below the strong coupling TCP as a function of the baryon chemical potential. Since at strong coupling, the nuclear critical point and the chiral TCP are on top, we show both fermionic observables for a better comparison. 
In Fig.~\ref{Fig:Susceptibility} we show both the baryon density and chiral susceptibilities for the same parameters, but now the peaks indicate the $\beta$-dependence of the transitions, which is expected to be first order (as indicated by finite size scaling). As the volume shown is rather small, the peaks are at a slightly larger value compared to the thermodynamic limit extrapolation at strong coupling. By comparing both orders, it seems that the shift in the peak towards smaller $\mu_B$ is consistent with increasing $\beta$ in both observables.
The average plaquette shown in Fig.~\ref{Fig:Plaquette} has a stronger $\beta$-dependence for $\Ord{\beta^2}$, which is clearly expected for a gauge observable as the plaquette occupation numbers are larger in this case compared to $\Ord{\beta}$. However, this observable has no imprint of the chiral transition. 


The result concerning the chiral transition in the phase diagram, comparing $\Ord{\beta}$ and $\Ord{\beta^2}$ will be discussed in detail in a forthcoming publication where more data at lower $\mu_B$ enter the analysis. But it is already evident from the results discussed here in the vicinity of the tri-critical point that both orders have the same $\beta$-dependence for fermionic observables and that the tri-critical point is almost invariant in the regime up to $\beta=1$.

\section{Summary and Outlook}
 
We have presented how the dual representation based on the the strong coupling expansion -- that is in principle valid to all orders in the inverse gauge coupling $\beta$ -- can be mapped on a vertex model for actual simulations in 3+1 dimensions. 
The caveat of this method is that it re-introduces the sign problem gradually with $\beta$. We truncated the strong coupling expansion at $\Ord{\beta^2}$ up to $\beta\lesssim 1$ to keep the Monte Carlo simulations feasible. 
We have analyzed a set of thermodynamic observables that are sensitive to the chiral and nuclear transition, comparing both $\Ord{\beta}$ and $\Ord{\beta^2}$ results in the vicinity of the tri-critical  point.
The tri-critical point has, a very weak $\beta$-dependence based on rather small volumes. If the tri-critical point remains invariant for higher orders in $\beta$, the critical point might also exist in the continuum.

In the regime up to $\beta=1$, higher orders of the strong coupling expansion would certainly not yield different results and are thus not necessary to pursue. Only for $\beta>1$ plaquette occupation numbers with $n_p, \bar{n}_p>2$ are necessary
and might indicate a stronger dependence of the transition on $\beta$. To overcome the severity of the sign problem, one either has to work out the resummations that would give rise to the character expansion, or resort to quantum simulations as they do not care about the sign problem. In the strong coupling limit, simulations on a quantum annealer have already been carried out 
\cite{Annealer:2023}, and for gate-based quantum computing, the set of gates for a correpsonding Hamiltonian have been worked out \cite{Fromm:2023}.

\acknowledgments
We thank Giuseppe Gagliardi who provided the program to compute the tensors from which we derived the vertex model used in the Monte Carlo simulations. 
The authors gratefully acknowledge the funding of this project by computing time provided
by the Paderborn Center for Parallel Computing (PC$^2$). 
This work is supported by the Deutsche Forschungsgemeinschaft (DFG) through the CRC-TR 211 ’Strong-interaction matter under extreme conditions’– project number 315477589 – TRR 211.


\clearpage


\begin{thebibliography}{99}

\bibitem{Borsanyi:2021}
S.~Borsanyi, Z.~Fodor, M.~Giordano, S.~D.~Katz, D.~Nogradi, A.~Pasztor and C.~H.~Wong,\\
\emph{Lattice simulations of the QCD chiral transition at real baryon density}.\\
Phys. Rev. D \textbf{105} (2022) no.5, L051506.
{\tt[arXiv:2108.09213 [hep-lat]]}.
\vspace{-2.5mm}
\bibitem{Bollweg:2022}
D.~Bollweg \textit{et al.} [HotQCD],\\
\emph{Taylor expansions and Pad\'e approximants for cumulants of conserved charge fluctuations at nonvanishing chemical potentials}.\\
Phys. Rev. D \textbf{105} (2022) no.7, 074511, {\tt[arXiv:2202.09184 [hep-lat]]}.
\vspace{-2.5mm}
\bibitem{Clarke:2024}
D.~A.~Clarke, P.~Dimopoulos, F.~Di Renzo, J.~Goswami, C.~Schmidt, S.~Singh and K.~Zambello,\\
\emph{Searching for the QCD critical endpoint using multi-point Pad\'e approximations}.\\
{\tt[arXiv:2405.10196 [hep-lat]]}.
\vspace{-2.5mm}
%
\bibitem{Hansen:2024}
M.~W.~Hansen and D.~Sexty,\\
\emph{Testing dynamical stabilization of Complex Langevin simulations of QCD}.\\
{\tt[arXiv:2405.20709 [hep-lat]].}
\vspace{-2.5mm}
%
\bibitem{Rossi:1984}
Pietro Rossi and Ulli Wolff,\\
\emph{Lattice {QCD} With Fermions at Strong Coupling: A Dimer System}.\\
 Nucl. Phys. B \textbf{248} (1984) 105.
%
\vspace{-2.5mm}
\bibitem{Gagliardi:2019}
G.~Gagliardi and W.~Unger.\\
\emph{New dual representation for staggered lattice QCD}.\\
Phys. Rev. D, \textbf{101} (2020) no.3, 034509.
{\tt[arXiv:1911.08389 [hep-lat]]}.
%
\vspace{-2.5mm}\bibitem{deForcrand:2014}
Ph.~de~Forcrand, J.~Langelage, O.~Philipsen, and W.~Unger.\\
\emph{Lattice QCD Phase Diagram In and Away from the Strong Coupling Limit}.\\
Phys. Rev. Lett., \textbf{113} (2014) no.15, 152002.
{\tt[arXiv:1406.4397 [hep-lat]]}.
%
\vspace{-2.5mm}
\bibitem{Kim:2023}
J.~Kim, P.~Pattanaik and W.~Unger,\\
\emph{Nuclear liquid-gas transition in the strong coupling regime of lattice QCD}.\\
Phys. Rev. D \textbf{107} (2023) no.9, 094514.
{\tt[arXiv:2303.01467 [hep-lat]]}.
\vspace{-2.5mm}
\bibitem{Kim:2016}
J.~Kim and W.~Unger,\\
\emph{Quark Mass Dependence of the QCD Critical End Point in the Strong Coupling Limit}.\\
PoS \textbf{LATTICE2016} (2016), 035.
{\tt[arXiv:1611.09120 [hep-lat]]}.
%
\vspace{-2.5mm}
\bibitem{Unger:2024}
W.~Unger,\\
\emph{The phase diagram at finite baryon and isospin densities at strong coupling}.\\
PoS \textbf{LATTICE2023} (2024), 176.
%
\vspace{-2.5mm}
\bibitem{Wenger:2008}
U.~Wenger,\\
\emph{Efficient simulation of relativistic fermions via vertex models}.\\
Phys. Rev. D \textbf{80} (2009), 071503.
{\tt[arXiv:0812.3565 [hep-lat]]}.
%
\vspace{-2.5mm}
\bibitem{Pattanaik:2023}
J.~Kim, P.~Pattanaik and W.~Unger,\\
\emph{Chiral transition via Strong Coupling expansion}.\\
PoS \textbf{LATTICE2023} (2024), 197
{\tt [arXiv:2312.14540 [hep-lat]]}.
%
\vspace{-2.5mm}
\bibitem{Bloch:2022}
J.~Bloch and R.~Lohmayer,\\
``Grassmann higher-order tensor renormalization group approach for two-dimensional strong-coupling QCD,''\\
Nucl. Phys. B \textbf{986} (2023), 116032
{\tt [arXiv:2206.00545 [hep-lat]]}.
%
\vspace{-2.5mm}
\bibitem{Langelage:2014}
J.~Langelage, M.~Neuman and O.~Philipsen,\\
\emph{Heavy dense QCD and nuclear matter from an effective lattice theory}.\\
JHEP \textbf{09} (2014), 131
{ \tt[arXiv:1403.4162 [hep-lat]]}.
%
\vspace{-2.5mm}
\bibitem{Annealer:2023}
J.~Kim, T.~Luu and W.~Unger,\\
\emph{U(N) gauge theory in the strong coupling limit on a quantum annealer}.\\
Phys. Rev. D \textbf{108} (2023) no.7, 074501
{ \tt[arXiv:2305.18179 [hep-lat]]}.
%
\vspace{-2.5mm}
\bibitem{Fromm:2023}
M.~Fromm, O.~Philipsen, W.~Unger and C.~Winterowd,\\
\emph{Quantum gate sets for lattice QCD in the strong-coupling limit: $N_{f}=1$}.\\
EPJ Quant. Technol. \textbf{11} (2024) no.1, 24
{\tt [arXiv:2308.03196 [hep-lat]]}.

\end{thebibliography}
\end{document}